\documentclass[12pt]{article}
\usepackage[utf8]{inputenc}
\usepackage[english]{babel}
\usepackage{csquotes}
\usepackage{comment}
\usepackage{xcolor}
\usepackage{appendix}
\usepackage{latexsym}
\usepackage{a4,amssymb, amsthm, amsmath}
\newtheorem{theorem}{Theorem}

\newtheorem{proposition}{Proposition}
\newtheorem{definition}{Definition}
\newtheorem{remark}{Remark}

\begin{document}
\title{Geometric theory of non-regular separation of variables and the bi-Helmholtz equation}

\author{Claudia Chanu\\
Dipartimento di Matematica, Universit\`a di Torino, \\
Torino, Italia \\
claudiamaria.chanu@unito.it
\and
Basel Jayyusi \\ 
Department of Applied Mathematics, University of Waterloo, \\
Waterloo,  Ontario, Canada \\
bjayyusi@uwaterloo.ca
\and
Raymond G McLenaghan \\
Department of Applied Mathematics, University of Waterloo, \\
Waterloo,  Ontario, Canada \\
rgmclena@uwaterloo.ca}
%\and
%email: claudiamaria.chanu@unito.it bjayyusi@uwaterloo.ca rgmclena@uwaterloo.ca}

\date{\today }

\maketitle
%\ArticleDates{Received ???, in final form ????; Published online ????}

\begin{abstract}{The geometric theory of additive separation of variables is applied to the search for multiplicative separated solutions of the bi-Helmholtz equation. It is shown that the equation does not admit regular separation in any coordinate system in any pseudo-Riemannian space. The equation is studied in the four coordinate systems in the Euclidean plane where the Helmholtz equation and hence the bi-Helmholtz equation is separable.  It is shown that the bi-Helmoltz equation admits non-trivial non-regular separation in both Cartesian and polar coordinates, while it possesses only trivial separability in parabolic and elliptic-hyperbolic coordinates.  The results are applied to the study of small vibrations of a thin solid circular plate of uniform density which is governed by the bi-Helmholtz equation.}
\end{abstract}
%\Keywords{variable separation; Hamilton-Jacobi equation; Killing tensors; (pseudo-)Riemannian manifolds}
%Please type here List of Keywords for your article separated by semicolon.

%\Classification{70H20, 70G45} % e.g. 35A30; 81Q05
% For 2000 Mathematics Subject Classification see http://www.ams.org/msc/

\section{Introduction}
The solution of boundary value problems for the partial differential equations of mathematical physics by the method of separation of variables is an effective method that has been employed for almost two centuries \cite{Morse1953, Moon1961}.
%has been employed since 1833, when Lam\'e first solved the heat equation in Euclidean space with respect to elliptic coordinates. 
The method, which assumes that the solution is a product of functions each one of a single independent variable ({\it product ansatz}), reduces the partial differential equation to a set of ordinary differential equations with a corresponding set of separation constants. The application of the method to the various equations of mathematical physics leads to certain ordinary differential equations the solutions of which are studied in their own right in special function theory. The special functions defined by an equation arising from a given separable coordinate system define function spaces that yield the solution of the given boundary problem under general boundary values.
The theory of this method for a general second order linear PDE with variable coefficients under the product ansatz is well developed \cite{Benenti2002a,Kalnins2018, Mille1977} 
%\textcolor{green}{We should cite Miller book, Kalnins book,Kalnins Kress Miller book}. 
%\textcolor{blue}{Maybe it's enough to cite the Kalnins Kress Miller 2018 book.}
However, the theory for equations of higher order has only  more recently been studied 
\cite{Kalnins1983b} 
where general separation is considered and where the concept of {\it non-regular separation} may be utilized.
In this article a geometric formulation of this theory \cite{Chanu2006a} is applied to the bi-Helmholtz equation which arises in the study of small vibrations of a thin, solid plate \cite{Rayleigh1877,Duff1966}.

The plan of the paper is as follows. In Section 2 the geometric theory of separation of variables is reviewed and the concepts of regular and non-regular separation defined. In Section 3 the theory of the previous section is applied to the bi-Helmholtz equation. The main result of this section is the proof of Theorem 1 which states that {\it regular multiplicative separation for the bi-Helmholtz equation $(\ref{eq_bilap})$ on any Riemannian or pseudo-Riemannian $n$-dimensional manifold  does not occur in any  system of coordinates.} Conditions for non-regular separation to occur are given in Proposition 2 and the concept of non-trivial non-regular separation is introduced. Section 4 contains examples in $\mathbb{E}^2$ of coordinate systems where the bi-Helmholtz equation admits trivial and non-trivial non-regular separation.  In Section 5 the results of the preceding section are applied to the analysis of the small vibrations of a thin solid circular plate which reproduce the classical results obtained by Rayleigh \cite{Rayleigh1877}. The Conclusion is given in Section 6.

\section{Geometric theory of separation of variables}
One of the most known ansatz for solving a PDE depending on $n$ independent variables $(q^i)$ is the additive separation, that is the search of those solutions $u(q^i)$ which are written
 as sum of functions depending on a single variable (in the given coordinate system $(q^i)$):
 $$
 u=\sum_{i=1}^n  S_{(i)}(q^i)
 $$
 Moreover, we do not look for a single separated solution but 
 we want to determine a family (as big as possible) of separated solutions
 $$
  u=\sum_{i=1}^n  S_{(i)}(q^i, c_\alpha),
 $$
 where the real constant parameters $c_\alpha$ satisfy a suitable completeness condition (see below). 
 When the method of SoV works, it is possible to split the PDE into $n$ ODEs of order $l$ which involve only one of the functions $S_{(i)}$; these ODEs are known as \emph{separated equations}.
 Separability of a PDE strongly depends on the choice of the independent variables and is destroyed by a general transformation of the $(q^i)$.
However, we want to provide a geometrical interpretation of what we mean by solving a PDE through the ansatz of separation of variables. We will distinguish between two possible types of families of separated solutions: the first type depends on the maximal number of parameters (we refer to this case as "free" \cite{Benenti2002a} or "regular" \cite{Kalnins1983b} separation of variables) and the second type depending on less parameters. We refer to this case as "non regular separation" \cite{Kalnins1983b,Chanu2008}.

We consider the separability of the $l$-th order PDE
\begin{equation} \label{eq}
\mathcal H(q^i,u,u_i,\ldots,u_i^{(l)})=h  \qquad (h\in \mathbb R)	
\end{equation}
in the coordinates $(q^i)$ on the $n$-dimensional manifold $Q$. Here and in the following we denote
by $u=u(q^i)$ the unknown function and set $$u_i=\frac{\partial u}{\partial q^i }, \qquad u^{(2)}_i=u_{ii}=\frac{\partial^2 u }{(\partial q^i)^2},\quad \ldots \quad u_i^{(l)}=\frac{\partial^l u}{(\partial q^i)^l}.$$ 
For the sake of simplicity, we assume that the maximal order of derivatives involved in $H$ is the same for each index $i=1,\ldots, n$.
Let $Z$ be the $(nl+1)$-dimensional space of the dependent variable and its separated derivatives: coordinates on $Z$ are given by
$(u, u_i, u_i^{(2)}, \ldots, u_i^{(l)})$. We consider the trivial bundle over $Q$, $M=Q\times Z$.
In \cite{Benenti2002a} {\it free separation} is defined as the existence of an additively separated solution $u$ of (\ref{eq}), depending on $nl+1$ parameters $(c_A)$, satisfying the completeness condition
	\[
\mathrm{rank}
\left[\frac{\partial u}{\partial c_A}\bigg | \frac{\partial u_i}{\partial c_A} 
\bigg | \ldots \bigg |\frac{\partial u_i^{(l)}}{\partial c_A}\right] = nl+1.	
\]
Free separation occurs if and only if 
 the $n$ vector fields of the form 
\begin{equation} \label{Di}
D_i=\frac\partial{\partial q^i}+u_i\frac\partial{\partial u}+u_i^{(2)}\frac\partial{\partial u_i}+\ldots
+u_i^{(l)}\frac\partial{\partial u_i^{(l-1)}}
+ R_i \frac\partial{\partial u_i^{(l)}},
\end{equation}
where $R_i(q^j,u,u_j,\ldots,u_j^{(l)})$ are functions on $M$,
are commuting symmetries of (\ref{eq}) namely, the conditions 
\begin{eqnarray} \label{tang}
	D_i\mathcal H=0, \\ \label{commu}
	[D_i,D_j]=0 
\end{eqnarray}
are satisfied (see \cite{Benenti2002a}). 
Eqs (\ref{tang}) determine the functions $R_i$, while (\ref{commu})  are equivalent to 
\begin{equation}\label{com}
D_iR_j=0 \qquad ( i\neq j).
\end{equation}
By expanding conditions (\ref{com}), we get exactly the conditions given in \cite{Kalnins1983b} for {\em regular separation}.
Thus, free separation corresponds to regular separation.

\begin{remark} \rm \label{r:grs}
The geometrical interpretation of the free separation is summarized in the following items (see \cite{Benenti2002a}):
\begin{itemize} 
\item
$\Delta=\hbox{\rm span}( D_i)$  is an integrable distribution of rank $n$ on $M$. 
\item 
The foliation of $n$-dimensional integral manifolds of $\Delta$ is described by
a {\em complete separated solution} of $\mathcal H=h$:
$$
\left\{ 
\begin{array}{l}
u=S=\sum_{i=1}^n  S_{(i)}(q^i, c_\alpha) \\
u_i=S_{(i)}^{\prime}\\
u_{ii}=S_{(i)}^{\prime\prime }\\
\cdots
\end{array}
\right.
\quad \hbox{with } \partial_i S=S_{(i)}^{\prime},\ \partial^2_{ij}S=\delta_{ij}S_{(i)}^{\prime\prime },\ \ldots 
$$
\item Free (regular) separated solutions depend on $nl+1$ parameters. 
\item 
Completeness means that for any point
$P_0$ with coordinates $(q_0^i)$ there is a separated solution of $\mathcal H=h$ for each choice of the value of $u$ and its $nl$ derivatives 
$u_i,u_i^{2},\ldots, u_i^{l}$ at $P_0$.
\end{itemize}
\end{remark}

\begin{remark} \label{rem-const} \rm
If $\mathcal{H}$ does not depend on $u$, then $u$ is defined up to an additive constant and the relevant constants appearing in the separated solutions are in fact $nl$ (the remaining one being the finial additive constant).
In this case we can eliminate the variable $u$ from the space $Z$ and  (\ref{Di}) reduce to
\begin{equation}\label{Dir}
    D_i=\frac\partial{\partial q^i}+u_i^{(2)}\frac\partial{\partial u_i}+\ldots
+u_i^{(l)}\frac\partial{\partial u_i^{(l-1)}}
+ R_i \frac\partial{\partial u_i^{(l)}}.
\end{equation}
Moreover, even the separated equations do not depend on $S_{(i)}$. Thus, each of $S_{(i)}$ is defined by its ODE up to an additive constant which can be disregarded, since it only affect the inessential global additive constant of $u$ . Hence, each $l$-th order ODE actually contribute with $l-1$ 
essential integration constants. This means that the $nl$ constants which are involved in the regular separated solutions splits into $n$ separation constants plus $nl-n=n(l-1)$ integrating constants (see Remarks 4.8 and 4.9 in \cite{Benenti2002a} for the case of the Schr\"odinger equation).   
\end{remark}
We consider now the second type of separation, occurring when condition (\ref{com}) is not identically satisfied. This case was firstly considered by 
Kalnins and Miller in \cite{Kalnins1983b}
and called {\em non-regular separation}. The authors state that also in this case separable solutions 
still may exist, but they will depend on less than $nl+1$  parameters. 

However, in \cite{Kalnins1983b} it is not specified under what conditions separable solutions will exist and how to determine the number of the  parameters involved.

In \cite{Chanu2008} one of us gives a geometric interpretation of the situation that naturally leads to an effective definition
of non-regular separation. 
%\textcolor{green}{I CHANGED THE NAME OF THE SUBMANIFOLD FROM $S$ TO $N$, SINCE $S$ IS ALREADY THE NAME OF THE SEPARATED SOLUTION}

Let $N$ be a submanifold of $M=Q\times Z$ locally described by the $r$ equations
$$
f_a=0 \qquad (a=1,\ldots,r). 
$$
If the vectors $D_i$ commute on $N$, that is 
$$
D_iR_j|_N=0 \qquad (i\neq j, \ i,j=1,\ldots,n),
$$
and the vector fields $D_i$ are tangent to $N$, that is 
$$
D_i f_a|_N=0 \quad (i=1,\ldots,n,\ a=1,\ldots r),
$$
then we can restrict the distribution $\Delta$ generated by  the $D_i$ to an involutive distribution on $N$. Thus,  
on $N$ we have a complete separated solution of $\mathcal H=h$ depending on $nl+1-r$ parameters $(c_\alpha)$,
which can be considered as a {\em constrained separated}
solution  on $M$. 
Hence, according to \cite{Chanu2008} we have the following

\begin{definition} \rm
The PDE $\mathcal H=h$ admits a {\em non-regular} or  {\emph constrained} additive separation on a submanifold $N$ of $M$
defined by the $r$ equations  $f_a=0$, if  
 \begin{enumerate}
\item $u=\sum_i S^{(i)}(q^i,c_\alpha)$ is a solution  of $\mathcal H=h$; 
\item $u$ depends on $nl+1-r$ parameters $(c_\alpha)$ satisfying the completeness conditions
$$
\hbox{rank} \left[\frac{\partial u}{\partial c_\alpha}\bigg | \frac{\partial u_i}{\partial c_\alpha} 
\bigg | \ldots \bigg |\frac{\partial u_i^{(l)}}{\partial c_A}\right] = nl+1-r;
$$
\item $u$ and its derivatives satisfy $f_\alpha(q^j,u,u_i...)=0$ for all (admissible) values of the parameters $(c_\alpha)$.
\end{enumerate}
\end{definition}

From the above discussion we get the following criterion for the constrained separation

\begin{proposition} \label{p:crit}
In a given coordinate system $(q^i)$ equation $(\ref{eq})$ admits a constrained separation on the submanifold $N$ defined by equations $f_a=0$ if and only if 
the vector fields $D_i$ $(\ref{Di})$ are  symmetries of $(\ref{eq})$, tangent to $N$ and commute on $N$, that is
$$
D_i \mathcal H=0, \quad D_iR_j|_N=0, \quad D_if_a|_N=0 \quad (i\neq j=1,\ldots, n, \; a=1,\ldots, r).
$$
\end{proposition}
\begin{remark} \rm
If $D_iR_j$ are everywhere different from zero, then equation (\ref{eq}) has no additive separable solutions.
A possible choice for  $N$ is the set of points satisfying equations 
$D_i R_j=0$, but in many cases this set is not a well-defined manifold or the vectors
$D_i$ are not tangent to it. 
\end{remark}

\begin{remark} \rm
In analogy with Remark \ref{r:grs}, the geometric interpretation of the constrained separation on a submanifold $N$ 
is sketched in the following items:
\begin{itemize} 
\item
$\Delta_S=\hbox{\rm span}( D_i)$ is an integrable $n$-dimensional distribution on $N$
whose integral manifolds are described by
a constrained complete separated solution of $\mathcal H=h$.
\item constrained separated solutions depend on $nl+1-r$ parameters.
\item completeness means that for any 
$P_0=(q^i_0)$ there is a separated solution of $\mathcal H=h$ for each choice of the value of $u$ and its $nl$ separated derivatives at 
$P_0$ satisfying $f_\alpha=0$ (i.e., the initial condition belongs to $N$).
\end{itemize}
\end{remark}

\section{Application to the bi-Helmholtz equation}
In order to apply the theory recalled in the previous section to the multiplicative separation of the bi-Helmholtz equation, we should transform the unknown function $\psi$ to its logarithm $u=\log \psi$ and then we should write down the simplified equation for separated solutions. First of of all, we need
 to express the bi-Helmholtz equation
\begin{equation}
\label{eq_bilap}
\Delta^2\psi=\lambda\psi, \qquad \lambda \in \mathbb R
\end{equation}
where $\Delta$ is the Laplace-Beltrami operator on a $n$-dimensional Riemannian or pseudo-Riemannian manifold $(Q,\mathbf g)$ with respect to a coordinate system $(q^i)$ in which the metric has contravariant components  $(g^{ij})$. 
For the moment we do not require any special property to the coordinates (such as orthogonality, or specifying of the dimension $n$).
It is well known that 
\begin{equation}
\label{eq_lap}
\Delta\psi=g^{ij}\partial_i\partial_j\psi -\Gamma^i \partial_i \psi
\end{equation}
where $\partial_i$ is the partial derivative with respect to $q^i$ and
$$
\Gamma^h=g^{ij}\Gamma_{ij}^h=\frac 12\, g^{ij}g^{hk}(\partial_i g_{jk}+\partial_jg_{ik}-\partial_kg_{ij})
$$
are the contracted Christoffel symbols (see \cite{Benenti2002a}). 

If not explicitly written, Einstein summation notation is sytematically employed: repeated upper and lower indices means summation. 

By applying twice the Laplace operator we get
\begin{eqnarray}
\Delta^2\psi&=&g^{ij}\partial_{ij}(g^{hk}\partial_{hk}\psi -\Gamma^h \partial_h \psi) -\Gamma^i \partial_i (g^{hk}\partial_{hk}\psi -\Gamma^h \partial_h \psi)= \nonumber
\\ \nonumber
&=& g^{ij}g^{hk}\partial_{ijhk}\psi + 2(g^{ij}\partial_jg^{hk}-g^{hk}\Gamma^i)\partial_{ihk}\psi +
\\
& & (g^{ij}\partial_{ij}g^{hk}-2g^{jk}\partial_j\Gamma^h-\Gamma^i\partial_i g^{hk}+\Gamma^h\Gamma^k)\partial_{hk}\psi+
\\ \nonumber
& & (-g^{ij}\partial_{ij}\Gamma^h +\Gamma^i\partial_i\Gamma^h) \partial_h \psi,
\end{eqnarray}
where
\begin{equation}
    \partial_{i_{1}\ldots i_{p}}:=\partial_{i_1}\ldots\partial_{i_{p}}.
\end{equation}

In order to simplify notation we can write the bi-Laplacian as
\begin{equation}
\label{eq_bilap_form}
\Delta^2\psi=A^{ijkl}\partial_{ijkl}\psi + B^{ijk}\partial_{ijk}\psi+C^{ij}\partial_{ij}\psi+D^i\partial_i \psi,
\end{equation}
where
\begin{eqnarray}
&&A^{ijkl}=g^{(ij}g^{kl)}
\\
&&B^{ijk}=2(g^{h(i}\partial_hg^{jk)}-g^{(ij}\Gamma^{k)}) 
\\
&&C^{ij}=g^{kl}\partial_{kl}g^{ij}-2g^{k(i}\partial_k\Gamma^{j)}-\Gamma^{k}\partial_{k} g^{ij}+\Gamma^i\Gamma^j
\\
&&D^i=-g^{jk}\partial_{jk}\Gamma^i +\Gamma^j\partial_j\Gamma^i=-\Delta \Gamma^i
\end{eqnarray}
%By the symmetry of the metric tensor it follows 
%$$
%A^{ihk}=A^{ikh}
%$$
%while the $B^{ij}$ are not symmetric (maybe they are if we insert %also the expressions of the $\Gamma^i$).
where $(\cdots)$ indicates symmetrization of the indices.

In order to pass from a multiplicative separated solution to an additive separated solution we perform the change of unknown
$$
\psi=e^u, \qquad u=\log \psi
$$
We denote the partial derivatives of $u$ by $u_i$, $u_{ij}$ etc. The link between partial derivatives of $\psi$ and $u$ is given by
\begin{equation*}
\begin{array}{l}
 \partial_i \psi=e^u u_i \vphantom{\dfrac 12}
\\
 \partial_{ij} \psi=e^u(u_{ij}+u_iu_j) \vphantom{\dfrac 12}
\\
 \partial_{ijk}\psi=e^u(u_{ijk}+ 3u_{(i} u_{jk)} + u_iu_ju_k) \vphantom{\dfrac 12}
\\ 
 \partial_{ijkl}\psi=
e^u(u_{ijkl}+ 4u_{(i} u_{jkl)} +
3u_{(ij}u_{kl)} +
6u_{(i}u_{j} u_{kl)} +
u_iu_ju_ku_l) \vphantom{\dfrac 12}
\end{array}
\end{equation*}
Hence, the equation to which we want to apply the geometric theory of SoV is
$$
H(q^i,u_i,u_{ij},\ldots,u_{ijkl})=\lambda, \qquad \lambda \in \mathbb R,
$$
where
$$
\begin{array}{l}
H=A^{ijkl}u_{ijkl}+(B^{ijk}+4A^{ijkl}u_l)u_{ijk}
+(C^{ij}+3B^{ijk}u_k)u_{ij}+  \\ \qquad
3A^{ijkl}u_{ij} u_{kl} + 6A^{ijkl}u_iu_ju_{kl}+ \vphantom{\dfrac 12}
\\ \qquad
A^{ijkl} u_i u_j u_k u_l + B^{ijk} u_i u_j u_k + C^{ij} u_i u_j+D^i u_i,
\end{array}
$$
If we are interested only in separated solutions $$u=\sum_iS_{(i)}(q^i,c_A), \qquad A=1,\ldots, 4n,$$ where, since the PDE does not depend explicitly on $u$, the constants $(c_A)$ are at most $4n$, we replace $H$ by 
\begin{equation}\label{eq_Hs}
\begin{array}{l}
H_s=(g^{ii})^2u_i^{(4)}+(4g^{ii}g^{ij}u_j+B^{iii})u_i^{(3)}+(g^{ii}g^{jj}+2(g^{ij})^2)u_i^{(2)}u_j^{(2)}
+ \\ \qquad \ 
(2(g^{ii}g^{hj}+2g^{ij}g^{ih})u_ju_h+(B^{iij}+B^{iji}+B^{jii})u_j+C^{ii})u_i^{(2)}+\vphantom{\dfrac 12} \\
\qquad \ g^{ij} g^{hk} u_i u_j u_h u_k + B^{ihk} u_i u_h u_k + C^{ij} u_i u_j+D^i u_i,
\end{array}
\end{equation}
where $u_i^{(s)}$ is the $s$-th partial derivative of $u$  w.r.t. $q^i$.

The function (\ref{eq_Hs}) is a 4-th degree polynomial in the derivatives of $u$ to the 4-th order; by observing that the terms $B^{ijh},C^{ij},D^i$ contains first, second, third order derivatives of the metric tensor respectively, we can say that $H_s$ is homogeneous: the sum of the degree and the order of the derivatives it is the same (four) in all terms.

For the further computation it is useful to write down the first partial derivatives of $H_s$
\begin{eqnarray}
\frac{\partial H_s}{\partial u^{(4)}_i}&=&(g^{ii})^2 \\
\frac{\partial H_s}{\partial u^{(3)}_i}&=&B^{iii}+4g^{ii}g^{ij}u_j \\
\frac{\partial H_s}{\partial u^{(2)}_i}&=&2(g^{ii}g^{jj}+2(g^{ij})^2)u_j^{(2)}+2(g^{ii}g^{hj}+2g^{ij}g^{ih})u_ju_h+ \\
& & (B^{iij}+B^{iji}+B^{jii})u_j+C^{ii} \nonumber \\
\frac{\partial H_s}{\partial u_i}&=& 4g^{jj}g^{ij}u_j^{(3)}+u_j^{(2)}(B^{jji}+B^{jij}+B^{ijj}+4(g^{jj}g^{hi}+2g^{ji}g^{jh})u_h)+\\
& & D^{i} +  (C^{ij}+C^{ji}) u_j + ( B^{ijh}+B^{jhi}+B^{hij}) u_j u_h + 4 g^{ij}g^{hk} u_j u_h u_k \nonumber
\end{eqnarray}

We compute the vector fields associated with the separation,
which are of the form (\ref{Dir}):
$$
D_i=\partial_i +u_i^{(2)}\frac{\partial}{\partial u_i} +u_i^{(3)}\frac{\partial}{\partial u_i^{(2)}}+u_i^{(4)}\frac{\partial}{\partial u_i^{(3)}}
+R_i\frac{\partial}{\partial u_i^{(4)}},
$$
where the index $i$ is not summed and the functions  $R_i(q^h,u_h,...u_h^{(4)})$ are determined by the condition 
$$
D_i(H_s)=0.
$$
Hence, under the technical assumption that $g^{ii}\neq 0$, we have
$$
R_i=-\left( \frac{\partial H_s}{\partial u_i^{(4)}}\right)^{-1}\left(\partial_i H_s +u_i^{(2)}\frac{\partial H_s}{\partial u_i} +u_i^{(3)}\frac{\partial H_s}{\partial u_i^{(2)}}+u_i^{(4)}\frac{\partial H_s}{\partial u_i^{(3)}}\right)=
$$
$$
=-\frac{1}{(g^{ii})^2}\left( \partial_i H_s + u_i^{(2)}\frac{\partial H_s}{\partial u_i} +u_i^{(3)}\frac{\partial H_s}{\partial u_i^{(2)}}+u_i^{(4)}\frac{\partial H_s}{\partial u_i^{(3)}}\right)
$$
The condition for free (or regular) separation is that
$$
D_iR_j=0 \qquad \forall i \neq j
$$
that is
$$
\partial_iR_j+u_i^{(2)}\frac{\partial R_j}{\partial u_i} +u_i^{(3)}\frac{\partial R_j}{\partial u_i^{(2)}}+u_i^{(4)}\frac{\partial R_i}{\partial u_i^{(3)}}+R_i\frac{\partial R_j}{\partial u_i^{(4)}}=0.
$$
We look for terms containing $u_h^{(3)}$ or $u_h^{(4)}$ and  such that the sum
of the orders of the derivatives of $u$ is six i.e., the terms containing  $u^{(4)}_h u^{(2)}_k$ or  $u^{(3)}_h u^{(3)}_k$ (for $k\neq h$) and with the components of the metric tensor not derived.
Since we have for $i\neq j$
$$
\frac{\partial R_j}{\partial u_i^{(4)}}=-\frac{1}{(g^{jj})^2}\left( \frac{\partial^2 H_s}{\partial q^j\partial u_i^{(4)}} +u_j^{(2)}\frac{\partial^2 H_s}{\partial u_i^{(4)}\partial u_j} +u_j^{(3)}\frac{\partial^2 H_s}{\partial u_i^{(4)}\partial u_j^{(2)}}+u_j^{(4)}\frac{\partial^2 H_s}{\partial u_j^{(3)}\partial u_i^{(4)}}\right)
$$
$$
=-\frac{\partial_j (g^{ii})^2}{(g^{jj})^2},
$$
the addendum $R_i\frac{\partial R_j}{\partial u_i^{(4)}}$ does not contain such a term, as well as $\partial_iR_j$.
Moreover, we have
$$
\frac{\partial R_j}{\partial u_i^{(3)}}=-\frac{1}{(g^{jj})^2}\left( \frac{\partial^2 H_s}{\partial q^j\partial u_i^{(3)}} +u_j^{(2)}\frac{\partial^2 H_s}{\partial u_i^{(3)}\partial u_j} +u_j^{(3)}\frac{\partial^2 H_s}{\partial u_i^{(3)}\partial u_j^{(2)}}+u_j^{(4)}\frac{\partial^2 H_s}{\partial u_j^{(3)}\partial u_i^{(3)}}\right)
$$
$$
=-\frac{1}{(g^{jj})^2}\left(\partial_j B^{iii}+4\partial_j(g^{ii}g^{ih})u_h+4u_j^{(2)}g^{ii}g^{ij}\right)
$$
Thus, in  the addendum $u^{(4)}_i \frac{\partial R_j}{\partial u_i^{(3)}},$ the function $D_iR_j$ contains the term
\begin{equation}\label{uj2ui4}
-4\frac{g^{ii}g^{ij}}{(g^{jj})^2}u_j^{(2)}u_i^{(4)}
\end{equation}
For the derivative of $R_j$ w.r.t. $u_i^{(2)}$ we have
$$
\frac{\partial R_j}{\partial u_i^{(2)}}=-\frac{1}{(g^{jj})^2}\left( \frac{\partial^2 H_s}{\partial q^j\partial u_i^{(2)}} +u_j^{(2)}\frac{\partial^2 H_s}{\partial u_i^{(2)}\partial u_j} +u_j^{(3)}\frac{\partial^2 H_s}{\partial u_i^{(2)}\partial u_j^{(2)}}+u_j^{(4)}\frac{\partial^2 H_s}{\partial u_j^{(3)}\partial u_i^{(2)}}\right)=
$$
$$
=-\frac{1}{(g^{jj})^2}\left( \frac{\partial^2 H_s}{\partial q^j\partial u_i^{(2)}}+u_j^{(2)}\frac{\partial^2 H_s}{\partial u_i^{(2)}\partial u_j} +2(g^{ii}g^{jj}+2(g^{ij})^2)u_j^{(3)} \right).
$$
Therefore, in  $u^{(3)}_i \frac{\partial R_j}{\partial u_i^{(2)}}$, the function $D_iR_j$ contains the term
\begin{equation}\label{uj3ui3}
\frac{-2}{(g^{jj})^2}(g^{ii}g^{jj}+2(g^{ij})^2)u_j^{(3)}u_i^{(3)}
\end{equation}
and, since $\frac{\partial H_s}{\partial u_i^{(2)}}$ does not depend on $u_h^{(3)}$ and $u_h^{(4)}$, we disregard other addenda of $\frac{\partial R_j}{\partial u_i^{(2)}}$.
The last term to consider is
$$
\frac{\partial R_j}{\partial u_i}=-\frac{1}{(g^{jj})^2}\left( \frac{\partial^2 H_s}{\partial q^j\partial u_i} +u_j^{(2)}\frac{\partial^2 H_s}{\partial u_i\partial u_j} +u_j^{(3)}\frac{\partial^2 H_s}{\partial u_i\partial u_j^{(2)}}+u_j^{(4)}\frac{\partial^2 H_s}{\partial u_j^{(3)}\partial u_i}\right)
$$
$$
=-\frac{1}{(g^{jj})^2}\left( 4g^{jj}g^{ij}u_j^{(4)}+ \textrm{terms with derivatives of $u$ of order less than $4^{th}$}\right)
$$
Thus $D_iR_j$ contains the term
\begin{equation}\label{uj4ui2}
-4\frac{g^{ij}}{g^{jj}}u_j^{(4)}u_i^{(2)}
\end{equation}
and no other terms in $u_h^{(4)}u_k^{(2)}$ or $u_h^{(3)}u_k^{(3)}$ are present in $D_iR_j$ except for (\ref{uj2ui4},\ref{uj3ui3},\ref{uj4ui2}).
It is easy to see that the three terms cannot be all zero since $g^{ii}\neq 0,$ for all $i$.
Hence, we have proved that
\begin{theorem}
Regular multiplicative separation for the bi-Helmholtz equation $(\ref{eq_bilap})$ on any Riemannian or pseudo-Riemannian $n$-dimensional manifold  does not occur in any  system of coordinates. 
\end{theorem}
\begin{remark} \rm
This means that we cannot find a family of 
separated solutions that depends on $4n$ parameters and such that the values of 
\begin{equation}\label{ui}
    u_i,\qquad u_{i}^{(2)}, \qquad u_{i}^{(3)}, \qquad u_{i}^{(4)}
\end{equation}
can be fixed in an arbitrary way. Going back to the 
multiplicatively separated solutions $\prod_i \psi_i$ of (\ref{eq_bilap}), the impossibility that regular separation occurs can be interpreted as follows: the one-to-one relationship existing  between (\ref{ui}) and
$$
\frac{\psi_i^\prime}{\psi_i} \qquad \frac{\psi_i^{\prime\prime}}{\psi_i} \qquad \frac{\psi_i^{\prime\prime\prime}}{\psi_i}
\qquad \frac{\psi_i^{\prime\prime\prime\prime}}{\psi_i}
$$
(see \cite{Benenti2002a} for the explicit link till order 2)
implies that it is not possible to assign
these values
in an arbitrary way at a point $q_0$ and finding a separated  solution satisfying these initial conditions.
\end{remark}

However, smaller families of separated solutions may exist. 
%it is an example in which 
In this case the theory of non-regular separation (see \cite{Chanu2008}) may be usefully applied. 

\begin{proposition}
In any coordinate system $(q^i)$ allowing regular separation for the Helmholtz equation, there exists a submanifold $N$ of  $M$ with dimension at least $2n$ where bi-Helmholtz admits non-regular separation.
\end{proposition}
The statement follows form the fact that every solution of the Helmholtz equation is also a  solution of the bi-Helmholtz equation. Hence, the $2n$-parameter families of the multiplicatively separated solutions of the Helmholtz equation are multiplicatively separated solution of
(\ref{eq_bilap}). Non-regular separation separation is said to be {\it non-trivial} if there exists a bigger family of separable solutions for the bi-Helmholtz equation than that for the Helmholtz equation. Otherwise, the separation is said to be {\it trivial}. A detailed analysis of some examples shows that non-trivial non-regular separation is possible.

\section{Examples}
In this section we provide some explicit computations of the manifolds $N$ allowing non-regular separation of variables for the bi-Helmholtz equation. For sake of simplicity, we restrict ourselves to the Euclidean plane. Moreover, in order to be sure that there exists a $N$ where non-regular separation occurs, we shall consider only coordinates that allow separation for the Helmholtz equation. 
Several situations will be described. In two of the four possible separable coordinate systems (Cartesian and polar) we show that non-trivial non-regular separation of the bi-Helmholtz equation is possible
%we are able to find a bigger family of multiplicative separated solutions for the bi-Helmholtz equation than the one obtained for the Helmholtz equation
(sections 4.1 and 4.2). One the other hand, if we use parabolic or elliptic-hyperbolic coordinates, the only  possibility is trivial separation.
%we show that all separated solutions of bi-Helmholtz equation already satisfy Helmholtz equation (section 4.3).

\subsection{Cartesian coordinates on the plane}

Let us examine the case of bi-Helmholtz equation on the plane in Cartesian coordinates $(q^1,q^2)=(x,y)$.
The only separability condition $D_1R_2=0$ becomes
\begin{equation}
-(2u^{(2)}_2u_2+u_2^{(3)})(2u^{(2)}_1u_1+u_1^{(3)})=0
\end{equation}
which is satisfied if one of the factors vanishes.
However, this condition alone does not define a submanifold such that $D_1$ and $D_2$ are tangent to it: for example
\begin{equation}
D_1(2u^{(2)}_1u_1+u_1^{(3)})=2(u_1^{(2)})^2+2u_1^{(3)}u_1+u_1^{(4)}
\end{equation}
Thus this condition must also be added in order to define a manifold where the non-regular separation could occur.
Let us call $N$ the submanifold defined by the equations
\begin{equation}
f_1= u_1^{(3)}+2u^{(2)}_1u_1=0,\qquad \qquad  f_2=u_1^{(4)}+2(u_1^{(2)})^2+2u_1^{(3)}u_1=0
\end{equation}
It is easy to check that $D_if_a|_N=0$ ($i=1,2$ $a=1,2$) and that $D_iR_j|_N=0$. Hence, on the 6-dimensional manifold $N$ non-regular separation occurs in Cartesian coordinates. The four derivatives of $S_2(y)$ ($u_2,u_2^{(2)},u_2^{(3)},u_2^{(4)}$) can be arbitrarily assigned at an initial point, while only two of the derivatives of $S_1(x)$ ($u_1,u_1^{(2)}$) are free and the remaining ones are determined by the equations $f_a$ defining $N$.

Since the equations for $N$ only involve the dependence of $u$ on $q^1=x$, they can be seen as separated equations for the function $S_1(q^1)$: in particular $f_1$ and $f_2$ (its differential consequence)
imply that $S_1$ has to satisfy
\begin{equation}\label{eqHelmCart}
    \frac{d^2}{dx^2}S_1+ (\frac{d}{dx}S_1)^2=c_1,\qquad c_1\in \mathbb{R} 
\end{equation}

It is interesting to remark that (\ref{eqHelmCart}) means precisely that $S_1$ is a separated solution of the Laplace equation. Indeed, by inserting the
function $\psi_1(x)=e^{S_1}$ in  (\ref{eqHelmCart}),  we get
$$
%\frac{d}{dx} \frac{\psi_1^{\prime\prime}}{\psi_1}=0 \qquad \Leftrightarrow %\qquad 
\psi_1^{\prime\prime}=c_1 \psi_1 \qquad (c_1 \in \mathbb{R}).
$$
%\textcolor{green}{The last expression can be read in terms of the function %$u_1$
%as $$
%u_{11}+u_1^2= c_1
%$$
%and the equations of $N$, i.e., $f_1=0$ and $f_2=0$, can be seen as the %differential conditions of it}.
However, the family of the separated solutions is bigger than the separated solutions of the Helmholtz equation, since for $S_2$ we can get functions which are not solutions of the Helmholtz equation but only of the bi-Helmholtz one.
Indeed, by (\ref{eqHelmCart}) the separated equation for $H_s$ written in Cartesian coordinates takes the form 
%no term containing $u_1$ and its derivative survive and
\begin{equation}\label{equ2cart}
        u_2^{(4)}+3(u_2^{(2)})^2 + 4 u_2 u_2^{(3)} +6(u_2)^2 u_2^{(2)}+(u_2)^4+2c_1\left(u_2^{(2)}+(u_2)^2\right) + c_1^2 - \lambda=0, 
\end{equation}
where $c_1=u_1^{(2)}+(u_1)^2$ which is constant on the constraint surface $N$. A simpler form for (\ref{equ2cart})
is given in terms of $\psi_2=\ln(S_2)$: with respect to this new independent variable the ODE becomes the linear 4-th order equation
$$
(\psi_2)^{(4)}+2 c_1 \psi_2'' +(c_1^2-\lambda) \psi_2=0.
$$

\subsection{Polar coordinates}

Let us examine the case of bi-Helmholtz equation on the plane in polar coordinates $(q^1,q^2)=(r,\theta)$.
The situation is similar, but is no longer symmetric in the two variables because we pass from two ignorable coordinates to one ignorable coordinate.
\begin{equation}\label{pol28}
(u_1^{(3)}r^2+2r(u_1)^2+u_1^{(2)}r-u_1+2u_1^{(2)}u_1 r^2)(2u^{(2)}_2u_2+u_2^{(3)})=0
\end{equation}
As in the Cartesian case, if we consider the condition on $\theta$-depending function, on the six-dimensional submanifold $N_{pol}$ defined by
\begin{equation}\label{pol29}
f_1=2u^{(2)}_2u_2+u_2^{(3)}=0, \qquad f_2=u_2^{(4)}+2(u_2^{(2)})^2+2u_2^{(3)}u_2=0
\end{equation}
reduced separation occurs.
Conditions (\ref{pol29}) mean that the following separated equation for $S_2(\theta)$ holds
\begin{equation}\label{eqHelmpol}
    \frac{d^2}{d\theta^2}S_2+ (\frac{d}{d\theta}S_2)^2=c_2,\qquad c_2\in \mathbb{R} 
\end{equation}
i.e., that 
$\psi_2(\theta)=e^{S_2}$ satisfies 
\begin{equation}\label{eq_theta}
    \psi_2^{\prime\prime}=c_2 \psi_2
\end{equation}
As in the Cartesian case plugging (\ref{eqHelmpol}) into the separated equation $H_s$ provides a separated equation which involves the variable $r$ only:
%INCLUDE SEPARATED EQUATIONS AND APPLICATION TO VIBRATING CIRCULAR PLATE.
\begin{align}
    &r^4u_1^{(4)}+\left(4r^4 u_1 +2r^3\right)u_1^{(3)}+3r^4 (u_1^{(2)})^2+r^2\left(6r^2(u_1)^2+6 r u_1 +2-1\right)u_1^{(2)}+ \nonumber\\
    &\!+r^4(u_1)^4+2r^3(u_1)^3 +(2c_2-1)(r^2(u_1)^2-r u_1)- \lambda r^4 +c_2^2 -4c_2 =0
\end{align}
here $c_2=u_2^{(2)}+(u_2)^2$ which is constant on the constraint surface $N_{pol}$.
The above equation becomes simpler passing to
$\psi_1=e^{S_1}$
\begin{equation}\label{eq_r}
    r^4(\psi_1)^{(4)}+2r^3(\psi_1)^{(3)}
+r(2c_2-1)(r\psi_1^{\prime\prime}-\psi_1^{\prime})+(c_2^2+4 c_2-\lambda r^4)\psi_1
=0
\end{equation}
whose solutions will be analyzed in Section 5.

The condition for the $r$-depending part of (\ref{pol28}) 
is equivalent to
$$
  r^2 (\frac{d^2}{dr^2}S_1+ (\frac{d}{dr}S_1)^2)-r \frac{d}{dr}S_1=c_1,\qquad c_1\in \mathbb{R} 
  $$
However, this condition and its differential conditions defines a submanifold which is tangent to the generator $D_1$ only for $\lambda=0$.
%$D_1
  
\subsection{Parabolic and Elliptic-Hyperbolic coordinates}

We consider the general case of Liouville coordinates on $\mathbb{R}^2$ which produces Cartesian, polar, parabolic and elliptic-hyperbolic coordinates as special cases when we demand that the Gaussian curvature vanishes. This coordinate system is characterized by the following metric 

\begin{align}
    g= (f(u) +g(v))(du^2 +dv^2),
\end{align}
for some arbitrary smooth functions $f$ and $g$. We first consider the Helmholtz equation
\begin{align}\label{bHeq}
    \Delta \psi = \gamma \psi,
\end{align}
where we recall that $\Delta$ denotes the Laplace-Beltrami operator \eqref{eq_lap}.  In Liouville coordinates it takes the form
\begin{align}\label{LBopL}
    \Delta \psi=\frac{1}{f+g} \left( \partial_u^2 \psi + \partial_v^2 \psi \right)
\end{align}
If we assume $\psi(u,v)=U(u)V(v)$ the Helmholtz equation may be written as 

\begin{align}
    \left(\frac{U''}{U}- \gamma f\right)+\left(\frac{V''}{V} - \gamma g \right) =0 
\end{align}

We observe that the equation separates for any smooth functions $f$ and $g$.

\begin{align}\label{SepLvl}
    U''-(c+\gamma f )U=0 \hspace{20mm} V'' + (c-\gamma g ) V =0, 
\end{align}
where $c$ denotes the separation constant.

We now turn our attention to the bi-Helmoltz equation. The Gaussian curvature in this coordinate system has the following form 

\begin{align}
    K =-\frac{1}{2(f+g)^2}\left( f'' + g'' - \frac{f'^2 + g'^2}{f+g}\right) 
\end{align}

From here on we will set $K=0$ in order to consider the parabolic and elliptic-hyperbolic cases in which we also have that $f'\neq 0, g'\neq 0 $. We use the Laplace-Beltrami \eqref{LBopL} to write the bi-Laplace operator as

\begin{align}
   \Delta^2 \psi=  \frac{\Delta_0^2 \psi}{(f+g)^2} - \frac{2}{(f+g)^3}( f'\partial_u \Delta_0\psi + g'\partial_v \Delta_0 \psi) +\frac{f''+g''}{(f+g)^3} \Delta_0\psi \label{LLwk=0}
\end{align}

From here we can write the bi-Helmholtz equation $\Delta^2 \psi = \lambda \psi$, using  \eqref{LLwk=0}. The product ansatz $\psi(u,v)=U(u)V(v)$ implies that \eqref{eq_bilap} takes the form 

\begin{align}\label{LvlSep}
    &f\frac{U^{(4)}}{U}+g \frac{V^{(4)}}{V} - 2f' \frac{U^{(3)}}{U}- 2g' \frac{V^{(3)}}{V} + f'' \frac{U''}{U}+g''\frac{V''}{V}  \nonumber \\
    +&\frac{V''}{V}\left(f''+2f\frac{U''}{U}-2f'\frac{U'}{U}\right) 
     +\frac{U''}{U}\left(g''+2g\frac{V''}{V}-2g'\frac{V'}{V}\right) \nonumber \\
     -&\lambda (f+g)^3=0
    %+& 2(f+g)\frac{U''V''}{UV} + g \frac{U^{(4)}}{U}+ f \frac{V^{(4)}}{V}- 2f' \frac{U'V''}{UV}-2g'\frac{V'U''}{UV} + f'' \frac{V''}{V}+g''\frac{U''}{U} \\
    %-& 3\lambda f^2 g -3 \lambda f g^2 =0 \nonumber
\end{align}
We observe that the above equation is not separable and hence does not admit regular multiplicative separation. This result is consistent with  the conclusion of Theorem 1. However, non-regular separation may be possible.  By taking derivatives of \eqref{LvlSep} we can obtain necessary conditions for separation (constraint equations). We find a first necessary condition by applying $\partial^2_{u,v}$ which yields 

\begin{align}\label{LvlCond1}
   f'\left(\frac{V^{(4)}}{V}\right)'+g'\left(\frac{U^{(4)}}{U}\right)'
   +\left(\frac{V''}{V}\right)'\left(f''+2f\frac{U''}{U}-2f'\frac{U'}{U}\right)' \nonumber \\
     +\left(\frac{U''}{U}\right)'\left(g''+2g\frac{V''}{V}-2g'\frac{V'}{V}\right)'
    - 6 \lambda f'g'(f+g)  =0 
\end{align}

We again note that this condition is not separable. We proceed by dividing by $f'g'$ and applying $\partial^2_{u,v}$ to get
the separable condition
\begin{align}\label{Lvlcond2}
    %\left( \frac{V^{(3)}}{g'V}- \frac{V''V'}{g'V^2}\right)'
    &\left(\frac{1}{g'}\left( \frac{V''}{V}\right)'\right)'
    \left(\frac{\left(f''+2f\frac{U''}{U}-2f'\frac{U'}{U}\right)'}{f'}\right)' \nonumber \\
    %+ \left( \frac{U^{(3)}}{f'U}- \frac{U''U'}{f'U^2}\right)'
    &+\left(\frac{1}{f'}\left( \frac{U''}{U}\right)'\right)'
    \left(\frac{\left(g''+2g\frac{V''}{V}-2g'\frac{V'}{V}\right)'}{g'}\right)'=0.
\end{align}

From here we obtain several cases: (i) neither
%$ \left( \frac{U^{(3)}}{f'U}- \frac{U''U'}{f'U^2}\right)'$ and $\left( \frac{V^{(3)}}{g'V}- \frac{V''V'}{g'V^2}\right)' $
$\left(\frac{1}{g'}\left( \frac{V''}{V}\right)'\right)'$ nor $\left(\frac{1}{g'}\left( \frac{V''}{V}\right)'\right)'$
vanish, from which we obtain $\lambda=0$  (ii) only one vanishes which also implies $\lambda=0$ (iii) both vanish which gives us a solution to the Helmholtz equation in Liouville coordinates. Hence, we do not get any non-trivial separated solutions. The proof in case (i) is given below and in Appendix A while that for cases (ii) and (iii) is given in  Appendix B.

Case(i): Dividing \eqref{Lvlcond2} by $\left(\frac{1}{f'}\left( \frac{U''}{U}\right)'\right)'$ $\left(\frac{1}{g'}\left( \frac{V''}{V}\right)'\right)'$
we obtain a separable equation.  Separating and integrating 
(see appendix A for this calculation) we find the following equations  
\begin{align}
    (2g-C)\frac{V''}{V}-2g'\frac{V'}{V}+g''-C_1g -C_2 &=0, \label{constraint1}\\
    (2f+C)\frac{U''}{U}-2f'\frac{U'}{U} +f''-D_1f-D_2&=0, \label{constraint1b}
\end{align}
where $C,C_1,C_2,D_1.D_2\in \mathbb{R}$. We can use these conditions in (\ref{LvlCond1}) to obtain
\begin{align}\label{SepDerivEq}
    \frac{1}{g'}\left(\frac{V^{(4)}}{V}\right)'+ \frac{1}{f'}\left(\frac{U^{(4)}}{U}\right)'  + \frac{D_1}{g'} \left(\frac{V''}{V}\right)' + \frac{C_1}{f'} \left(\frac{U''}{U}\right)'- 6 \lambda (f+g)=0.
\end{align}
Notice that this condition is now separable. And using this condition we can integrate back to get a simplified form of equation
\eqref{LvlSep} 
\begin{align}\label{SimpleSep}
    &f\frac{U^{(4)}}{U}+g \frac{V^{(4)}}{V} - 2f' \frac{U^{(3)}}{U}- 2g' \frac{V^{(3)}}{V} + f'' \frac{U''}{U}+g''\frac{V''}{V} - \lambda f^3 -\lambda g^3+ \nonumber \\
    &+f\frac{V^{(4)}}{V} +g\frac{U^{(4)}}{U}+ (C_1g+C_2)\frac{U''}{U} +(D_1f+D_2)\frac{V''}{V}-3\lambda f^2 g-3\lambda fg^2 =0.
\end{align}

To make sure we account for all the constraints properly we also need to use the derivatives of equations \eqref{constraint1} and \eqref{constraint1b}. There are other constraints coming from the additional requirement $K=0$, for the interest of brevity we will include these calculations Appendix A.

After using all the constraint equations we end up with the following 

\begin{align}
&\frac{(C_2-D_2-(C_1+D_1 )f)}{C_1+D_1}\left(\alpha f +\beta +3 \lambda f^2 - \frac{(k+C_1)(D_1-k)f+C_1(D_2-D)+D(D_1 - k)}{2f+C}  \right.\nonumber \\
&+2\frac{f'^2 }{(2f+C)^2}(D_1-k)-\left. \frac{(D_1f+D_2)^2-f''^2}{(2f+C)^2} \right) 
-2\frac{(D_1-k)}{2f+C}f'^2 \nonumber \\
&+ \left((k-C_1)f+C_2+D\right)\left(\frac{(D_1-k)f+D_2-D}{2f+C}\right)+2\lambda f^3 + \alpha f^2 +(\beta+\gamma)f= \delta
\end{align}

Where from the condition $K=0$ we have $f'^2=kf^2+2Df-k$ and  $g'^2=-kg^2 +2D g + k$. Thus the above equation simplifies to a polynomial in $f$ after multiplication by $(2f+C)^2$. 
Furthermore, since $f'\neq 0 $, the set $\{1, f, f^2, \ldots, f^n\}$ is linearly independent. 
%the powers of $f$ are linearly independent 
%since $W(f^n,f^m) = (n-m)f'f^{n+m-1}$ and  $f'\neq 0 $, 
This implies that the coefficients of the different powers of $f$ must all vanish. The coefficient of the highest power of $f$ the is $8\lambda$ the vanishing of which implies that $\lambda =0$. In the case that $C_1+D_1 =0$ \eqref{FODE1} is a polynomial in $f$ with highest order term $3\lambda f^2$ which also implies $\lambda=0$.

\section{
The circular vibrating plate
}

%\section{Vibrating circular clamped plate}

As mentioned in Section 3 one can pass between multiplicative and additive separation by the change of variable $\psi=e^u$. In this section we use the framework of multiplicative separation.

We apply the results for polar coordinates obtained in Section 4 to study vibrations of a thin, solid plate of uniform material, constant thickness, and areal density $\rho$ \cite{Rayleigh1877,Duff1966}.

The equation of motion for the plate under the assumption of small oscillations is given by \cite{Duff1966}
\begin{align}\label{BwaveEq}
     \rho \ddot{\psi}+c\Delta^2 \psi=0,
\end{align}
where $\psi$ is the lateral elevation at any point on the region $\Omega$ in the interior of the plate and
%at every point of the plate and 
%$c^4$ is some positive constant of the material which contains the density, Young's modulus and thickness.
$c$ is some positive constant depending on the material. If the plate is clamped at the edge he boundary conditions on $\psi $ are $\psi\lvert_{\partial \Omega}=\psi_n\lvert_{\partial \Omega}=0$, where the subscript $n$ denotes differentiation with respect to the normal to the boundary. For the case of a circular plate $\Omega$ is a disk of some radius $a$. We obtain the solution of \eqref{BwaveEq} by utilizing a separation of variables approach rather that the Fourier series method employed by Rayleigh \cite{Rayleigh1877}. 

Consider solutions of (\ref{BwaveEq}) of the form $\psi= w(r,\theta)T(t)$ which yield the separated equations

%On the region $\Omega$ in the interior of the plate, $\psi$ is the lateral elevation at every point of the plate and $c^4$ is some positive constant of the material which contains the density, Young's modulus and thickness. A derivation of this equation can be found in Rayleigh\cite{Rayleigh1877} or Duff \& Naylor\cite{Duff1966}. If the plate is clamped the boundary conditions on $u$ are $u\lvert_{\partial \Omega}=u_n\lvert_{\partial \Omega}=0$ where the subscript $n$ denotes differentiating with respect to the normal coordinate to the boundary. For the case of a circular plate we have that $\Omega$ is a disk of some radius $a$. First we separate out the time variable $\psi= w(r,\theta)T(t)$

\begin{align}\label{tsep}
    \frac{\Delta^2 w }{w} = -\frac{\rho}{c} \frac{\ddot{T}}{T}=k^4.
\end{align}
The separation constant is assumed to be positive (hence written as $k^4 $ for convenience), since we expect oscillatory behaviour in time.
%Where we used a positive separation constant since we expect oscillatory behaviour in time. 
Notice that this separation is regular since there are no additional constraints on the separated functions. The time part of (\ref{tsep}) has the form
%We define $\omega^2=\frac{ck^4}{\rho}$, so the time part will have the form 
\begin{align}\label{tpart}
 \ddot{T}+\omega^2T=0,   
\end{align}
where $\omega^2=\frac{ck^4}{\rho}$. 
The general solution is given by
\begin{align}\label{tgsoln}
    T(t)= G \cos(\omega t) +H\sin(\omega t), 
\end{align}
where $G$ and $H$ are arbitrary constants.

The spatial part (\ref{tsep}) satisfies the bi-Helmholtz equation, 

\begin{align}
    \Delta^2 w = k^4 w 
\end{align}
Written out explicitly in polar coordinates this equation takes the form

\begin{align}\label{PlatebiH}
w_{rrrr}+ \frac{1}{r^4} w_{\theta \theta \theta \theta} + \frac{2}{r}w_{rrr} - \frac{1}{r^2}w_{rr}+ \frac{1}{r^3 } w_r + \frac{4}{r^4}w_{\theta\theta }- \frac{2}{r^3}w_{r\theta \theta } + \frac{2}{r^2} w_{rr\theta \theta }=k^4 w,
\end{align}
where the subscript $r$ denotes $\frac{d}{dr}$.
We now consider product solutions of the form
\begin{align}\label{prodsoln}
    w(r,\theta)=R(r)\Theta(\theta),
\end{align}
which yields the separated equations
\begin{align}\label{sepeq1}
   \Theta''=-l^2\Theta,
\end{align}
\begin{align}\label{sepeq2}
  D^4R + \frac{2}{r} D^3R-\frac{1}{r^2}(1+2l^2)D^2R+\frac{1}{r^3}(1+2l^2)DR-\frac{l^2}{r^4}(4-l^2)R=k^4R,
\end{align}
%\begin{align}\label{sepeq2}
  %R^{(4)} + \frac{2}{r} R^{(3)}-\frac{1}{r^2}(1+2l^2)R''+\frac{1}{r^3}(1+2l^2)R'-\frac{l^2}{r^4}(4-l^2)R=k^4R,
%\end{align}
where $l$ is a separation constant and $D=\frac{d}{dr}$. 
%where $D=\frac{d}{dr}$.
%\begin{align}
    %\left(\frac{d^2}{dr^2}+\frac{1}{r}\frac{d}{dr}-\frac{n^2}{r^2}+k^2\right)\left(\frac{d^2}{dr^2}+\frac{1}{r}\frac{d}{dr}-\frac{n^2}{r^2}-k^2\right)R_n &=0 \\
    %\left(\frac{d^2}{dr^2}+\frac{1}{r}\frac{d}{dr}-\frac{n^2}{r^2}-k^2\right)\left(\frac{d^2}{dr^2}+\frac{1}{r}\frac{d}{dr}-\frac{n^2}{r^2}+k^2\right)R_n&=0
%\end{align}

The solution of the differential equation (\ref{sepeq1}) satisfied by the angular function $\Theta$ has the form
\begin{align}
    \Theta_l(\theta) = E_l \cos(l\theta) + F_l \sin(l\theta),
\end{align}
%$$\Theta_n = E_n \cos(n\theta) + F_n \sin(n \theta)$$
where $E_l$ and $F_l$ are arbitrary constants. In order that the  function $\Theta$ be single valued on $\Omega$, the constant $l$ must be a positive integer
$$l=n \quad (n=1, 2, \ldots).$$

We now turn our attention to (\ref{sepeq2}). It 
may be shown that the left-hand-side may be factored in two ways to yield
%For $\Theta$ to be single valued we require $l=n$, this makes $\Theta$ $2\pi$ periodic.
%Now to separate the spatial coordinates first we require our solution to be $2\pi$ periodic in $\theta$, this means that we can write $w_n=R_n(r)\Theta_n(\theta)$ where $\Theta_n''=-n^2 \Theta$, substituting this into the $(\ref{PlatebiH})$ we have the following equation 
\begin{align}
    \left(D^2+\frac{1}{r}D-\frac{n^2}{r^2}+k^2\right)\left(D^2+\frac{1}{r}D-\frac{n^2}{r^2}-k^2\right)R &=0, \\
    \left(D^2+\frac{1}{r}D-\frac{n^2}{r^2}-k^2\right)\left(D^2+\frac{1}{r}D-\frac{n^2}{r^2}+k^2\right)R &=0.
\end{align}
This result shows that the differential operators which appear inside the  brackets (the Bessel and modified Bessel operators) commute.
%We have explicitly written the equation in two ways to show that these operators commute. 
Since we know bases of the solution spaces to the Bessel and modified Bessel equations, 
%are given by the Bessel and modified Bessel functions $ J_n(kr), Y_n(kr), I_n(kr), K_n(kr),$ 
the general solution of the radial equation may be written as

\begin{align}
    R_n(r) &=  A_n J_n(kr) + B_n Y_n(kr)+C_n I_n(kr) +D_n K_n(kr),
\end{align}
where $J_n,Y_n,I_n,K_n$ are respectively the Bessel and modified Bessel functions of the first and second kind and $ A_n, B_n, C_n, D_n$ are arbitrary constants. 

The Bessel functions of the second kind are inadmissible since they are singular at the origin so we take $B_n=D_n=0$. %The angular function satisfies the second order DE and thus has the form

%One retrieves the solution to the vibrating membrane by setting $C_n=0$. 
The boundary conditions tell us that $R_n(a)=R_n'(a)=0$.

\begin{align}
    R_n(a) &= A_n J_n(ka) + C_n I_n (ka) =0 \label{Bd} \\
    R_n'(a) &=kA_n J_n'(ka) +kC_n I_n'(ka) =0 \label{Bddv}
\end{align}

To solve ($\ref{Bd}$) we either need $ka=j_{n,m}$ (the $m$th zero of $J_n$) and $C_n=0$ in which case we retrieve the vibrating membrane solution, or $A_n=-\frac{I_n(ka)}{J_n(ka)}C_n$ (we can absorb $C_n$ into the definition of $E_n,F_n$ to simplify). Since we are interested in solutions other than the ones for the vibrating membrane we take the latter in which case (\ref{Bddv}) becomes 

\begin{align}
    I_n'(ka)-\frac{I_n(ka)}{J_n(ka)}J_n'(ka)&=0 \\
    \frac{I_n'(ka)}{I_n(ka)}- \frac{J_n'(ka)}{J_n(ka)}&=0 \label{Besselcond}
\end{align}

We determine the roots of the above equation numerically to get a condition $ka=l_{n,m},$ where $l_{n,m}$ is the $m$th root of equation (\ref{Besselcond}) for some fixed $m$ 

\begin{align}
    R_{n,m}(r)=C_n\left(  I_n\left(\frac{l_{n,m}r}{a}\right)-\frac{I_n(l_{n,m})}{J_n(l_{n,m})}J_n\left(\frac{l_{n,m} r}{a}\right)\right)
\end{align}

Thus the general solution for $u$ has the form 

\begin{align}
    \psi=\sum\limits_{m,n} \left(E_n\cos(n\theta) +F_n\sin(n\theta)\right)\left(G_{n,m}\cos(\omega_{n,m}t) + H_{n,m}\sin(\omega_{n,m}t)\right)R_{n,m}
\end{align}

Where $\omega_{n,m}= c^2\frac{l_{n,m}^2}{a^2}$. See Rayleigh \cite{Rayleigh1877} for further study of this solution.

\section{Conclusion}
The main idea of the paper is to 
apply the technique of the  regular and non-regular separation to 
the search of the multiplicative separated solutions of the bi-Helmholtz equation, which is a classical 4-th order PDE equation of Mathematical Physics which was solved in particular cases by means of this technique (\cite{Rayleigh1877}). 
Nevertheless,  deeper studies about separability usually deal 
with first and second order PDE's (as well as the Hamilton-Jacobi equation and the stationary Schr\"odinger equation).
 
The choice of the bi-Helmholtz equation relies on two aspects: the existence of a physical application (it is not a simple toy model) and the fact that this kind of equation seemed to provide a good example for a deeper understanding of the non-regular separation method.

Non-regular separation appears naturally in the study of separability of the bi-Helmholtz equation, since we prove (Section 3) that regular separation never occurs for this equation, but on the other hand
we already know the existence of a family of separated solutions (the trivial ones, arising from the separability of the standard Helmholtz equation). In the examples on the Euclidean plane, we show that, in two of the four separable coordinate systems (Cartesian and polar coordinates), non-trivial separated solutions can be determined, while in the remaining  coordinate systems (parabolic and elliptic-hyperbolic coordinates) the only possibles separated solutions are proved to be the trivial ones.  

The study of the separated solutions is done from the simplest geometrical view point: the search for a submanifold, as large as possible, where the separability conditions are satisfied, without assuming any particular structure for the separated solution (such as as side conditions or generalizations of St\"ackel matrices as is done in \cite{Kalnins2018}). 

This paper  represents the first step in a program to analyse interesting examples of non-regular separation
As further extensions of the present paper, one could consider examples in higher dimensions or in different Riemannian manifolds, to get more physical applications.
%Moreover,
Indeed, a wider collection of examples is the natural starting point of a study to gain a deeper insight into the geometric conditions (for instance in terms of symmetry operators) of the non-regular separation of the bi-Helmholtz equation, as has been done 
%as instance for a well-known case of non-regular separation: 
for the fixed energy R-separation for the Schr\"odinger equation \cite{Chanu2006a, Kalnins2018}.

\section*{Acknowledgements} R. G. McLenaghan
wishes to acknowledge financial support from the Natural Sciences and Engineering Research Council of Canada in the form of and a Discovery Grant. He would also like to thank the Dipartimento di Matematica of University of Torino (Italy) for hospitality during several visits when part of this research was undertaken. C. M. Chanu  acknowledges the contribution of  the local research project {\it Metodi  Geometrici  in  Fisica  Matematica e Applicazioni (2019)} of Dipartimento di Matematica of University of Torino.

%\section*{Data Availability}
%Data sharing is not applicable to this article as no new data were created or analyzed in this
%study.

%\pagebreak

\appendix

\section{}

Dividing \eqref{LvlCond1} by $\left( \frac{U^{(3)}}{f'U}- \frac{U''U'}{g'U^2}\right)'\left( \frac{V^{(3)}}{g'V}- \frac{V''V'}{g'V^2}\right)' $, we see that we can separate as follows

\begin{align}
    \frac{\left(\frac{\left(g''+2g\frac{V''}{V}-2g'\frac{V'}{V}\right)'}{g'}\right)'}{\left( \frac{V^{(3)}}{g'V}- \frac{V''V'}{g'V^2}\right)'}= -\frac{\left(\frac{\left(f''+2f\frac{U''}{U}-2f'\frac{U'}{U}\right)'}{f'}\right)'}{\left( \frac{U^{(3)}}{f'U}- \frac{U''U'}{f'U^2}\right)'}=C
\end{align}

Where $C\in \mathbb{R}$. Separating and integrating 

\begin{align}
\frac{\left(g''+2g\frac{V''}{V}-2g'\frac{V'}{V}\right)'}{g'} &= \frac{C}{g'} \left(\frac{V''}{V}\right)'+C_1\\
\frac{\left(f''+2f\frac{U''}{U}-2f'\frac{U'}{U}\right)'}{f'} &= -\frac{C}{f'} \left(\frac{U''}{U}\right)'+D_1
\end{align}

The case where one of $\left( \frac{U^{(3)}}{f'U}- \frac{U''U'}{g'U^2}\right)' , \, \left( \frac{V^{(3)}}{g'V}- \frac{V''V'}{g'V^2}\right)' $ vanish corresponds to setting $C=0$ in one of the above equations. Multiplying through by $f'$ and $g'$ respectively and integrating once again we obtain

\begin{align}
g''+2g\frac{V''}{V}-2g'\frac{V'}{V}&= C \frac{V''}{V}+C_1g +C_2 \\
f''+2f\frac{U''}{U}-2f'\frac{U'}{U}&= -C \frac{U''}{U}+D_1 f +D_2
\end{align}

Where $C_1,C_2,D_1.D_2\in \mathbb{R}$. Or more compactly

\begin{align}
    (2g-C)\frac{V''}{V}-2g'\frac{V'}{V}+g''-C_1g -C_2 &=0 \label{constraint2}\\
    (2f+C)\frac{U''}{U}-2f'\frac{U'}{U} +f''-D_1f-D_2&=0 \label{constraint2b}
\end{align}

%\pagebreak

We can now separate \eqref{SepDerivEq}

\begin{align}\label{twosepeq}
    \frac{U^{(4)}}{U}+C_1\frac{U''}{U}-3\lambda f^2&=\alpha f+ \beta \\
    \frac{V^{(4)}}{V}+D_1\frac{V''}{V}-3\lambda g^2&=-\alpha g+ \gamma
\end{align}
 
The condition $K=0$ separates into 
\begin{align}
\frac{f^{(3)}}{f'}=-\frac{g^{(3)}}{g'}=k
\end{align}

Where $k\in\mathbb{R}$ is a separation constant. Integrating once 

\begin{align}
    f''=kf+D \hspace{10mm}g''=-kg+D
\end{align}

These are extra conditions that have to be taken into account. We also need to account for the derivatives of equation \eqref{constraint2} and \eqref{constraint2b}

\begin{align}\label{reducedsep}
    \frac{U''}{U}&= \frac{2f'}{2f+C} \frac{U'}{U} + \frac{D_1 f +D_2-f''}{2f+C}\\
    \frac{U^{(3)}}{U}&= \frac{D_1 f +D_2+f''}{2f+C} \frac{U'}{U}+ \frac{D_1-k}{2f+C}f'\\
    \frac{U^{(4)}}{U} &= \frac{D_1}{2f+C}f''+\frac{D_1 f +D_2+f''}{2f+C} \frac{U''}{U}- \frac{2f'}{2f+C}\frac{U^{(3)}}{U}\\
    &= \frac{(D_1-k)f''}{2f+C}+ \frac{(D_1f+D_2)^2-f''^2}{(2f+C)^2}- \frac{2f'^2}{(2f+C)^2}(D_1-k)\\
    \frac{V''}{V}&= \frac{2g'}{2g-C} \frac{V'}{V} + \frac{C_1 g +C_2+g''}{2g-C}\\
    \frac{V^{(3)}}{V}&= \frac{C_1 g +C_2+g''}{2g-C} \frac{V'}{V}+ \frac{C_1+k}{2g-C}g'\\
    \frac{V^{(4)}}{V} &= \frac{C_1}{2g-C}g''+\frac{C_1 g +C_2-g''}{2g-C} \frac{V''}{V}- \frac{2g'}{2g-C}\frac{V^{(3)}}{V}\\
    &= \frac{(C_1+k)g''}{2g+C}+ \frac{(C_1g+C_2)^2-g''^2}{(2g-C)^2}- \frac{2g'^2}{(2g-C)^2}(C_1+k)
\end{align}

\pagebreak

Using \eqref{reducedsep} to eliminate higher derivatives in \eqref{twosepeq}

\begin{align}\label{FODE1}
&(C_1+D_1) \frac{2f'}{2f+C} \frac{U'}{U} -2\frac{f'^2}{(2f+C)^2} (D_1-k)+(D_1-k)\frac{f''}{2f+C} +\frac{(D_1f+D_2)^2-f''^2}{(2f+C)^2} + \nonumber \\
+& C_1 \frac{D_1f+D_2-f''}{2f+C} - 3\lambda f^2 = \alpha f + \beta \\
&(C_1+D_1) \frac{2g'}{2g-C} \frac{V'}{V} -2\frac{g'^2}{(2g-C)^2} (C_1+k)+(C_1+k)\frac{g''}{2g-C} +\frac{(C_1g+C_2)^2-g''^2}{(2g-C)^2} + \nonumber \\
+& D_1 \frac{C_1g+C_2-g''}{2g-C} - 3\lambda g^2 = -\alpha g+ \gamma
\end{align}

 Substituting \eqref{twosepeq} into (\ref{SimpleSep}) we have the following

\begin{align}\label{E98}
-&2f'\frac{U^{(3)}}{U} - 2g' \frac{V^{(3)}}{V} + \left((k-C_1)f+C_2+D\right)\frac{U''}{U}+ \left(D_2+D-(k+D_1)g\right)\frac{V''}{V} \nonumber\\
    +&2\lambda (f^3+g^3) +\alpha(f^2-g^2)+(\beta+\gamma)(f+g)=0
\end{align}

Separating this equation we have 

\begin{align}\label{E100}
-2f'\frac{U^{(3)}}{U}+\left((k-C_1)f+C_2+D\right)\frac{U''}{U} + 2 \lambda f^3 + \alpha f^2 +(\beta+\gamma)f &= \delta \\
    -2g'\frac{V^{(3)}}{V}+\left(-(k+D_1)g+D_2+D\right)\frac{V''}{V} + 2 \lambda g^3 - \alpha g^2 +(\beta+\gamma)g &= -\delta
\end{align}

For some $\delta \in \mathbb R$. Eliminating the derivatives from $(\ref{E100})$ we have the following

\begin{align}\label{FODE2}
 &2f'\frac{C_2-D_2-(C_1+D_1)f}{2f+C}\frac{U'}{U}-2\frac{(D_1-k)}{2f+C}f'^2 \\
 &+\left((k-C_1)f+C_2+D\right) \left( \frac{(D_1-k)f+D_2-D}{2f+C}\right) \nonumber \\
    +&2\lambda f^3 + \alpha f^2 +(\beta+\gamma)f= \delta   
    \end{align}

Isolating for $\frac{2f'}{2f+C}\frac{U'}{U}$ from $(\ref{FODE1})$ assuming $C_1+D_1\neq 0$ and using this to eliminate derivatives in \eqref{FODE2}

\begin{align}
&\frac{(C_2-D_2-(C_1+D_1 )f)}{C_1+D_1}\left(\alpha f +\beta +3 \lambda f^2 - \frac{(k+C_1)(D_1-k)f+C_1(D_2-D)+D(D_1 - k)}{2f+C}  \right.\nonumber \\
&+2\frac{f'^2 }{(2f+C)^2}(D_1-k)-\left. \frac{(D_1f+D_2)^2-f''^2}{(2f+C)^2} \right) 
-2\frac{(D_1-k)}{2f+C}f'^2 \nonumber \\
&+ \left((k-C_1)f+C_2+D\right)\left(\frac{(D_1-k)f+D_2-D}{2f+C}\right)+2\lambda f^3 + \alpha f^2 +(\beta+\gamma)f= \delta
\end{align}

When $C_1 +D_1 =0$, \eqref{FODE1} implies $\lambda =0$.

\section{}

Case (ii): In this case we have 
\begin{align}
  \left(\frac{1}{f'} \left( \frac{U''}{U} \right)' \right)'=0,  \quad \left(\frac{1}{g'}\left( \frac{V''}{V}\right)'\right)'\neq 0.
\end{align}
Integrating we get 
\begin{align}\label{spc3}
   \frac{U''}{U}=\alpha_5 f +\alpha_6,
\end{align}
where $\alpha_5,\alpha_6$ are constants. Equation \eqref{Lvlcond2} implies that

\begin{align}\label{spc4}
    2f\frac{U''}{U}-2f'\frac{U'}{U}+f'' = \alpha_3 f+ \alpha_4
\end{align}

$\alpha_3,\alpha_4$ are constants. Substitution for $\frac{U''}{U}$ from \eqref{spc3} yields
\begin{align}\label{spc5}
    2f'U'=(f''+2\alpha_5 f^2+(2\alpha_6 -\alpha_3)f - \alpha_4)U.
\end{align}
Differentiation of the above equation followed substitution for $U''$ from \eqref{spc3} and $2f'U'$ from \eqref{spc5} yields after simplification
\begin{align}\label{spc5a}
    (2\alpha_5 f^2+(2\alpha_6-\alpha_3)f-\alpha_4^2)^2-f''^2+2f'f^{(3)}+4\alpha_5 ff'^2-\alpha_3 f'^2=0
\end{align}

Using the relations between $f$ and its derivatives

\begin{align}
f''=kf+D, \quad  f'^2=kf^2+2Df+\Lambda
\end{align}

Equation \eqref{spc5a} becomes a polynomial in $f$, the coefficient of the highest power of $f$ is $4\alpha_5^2$ which implies that $\alpha_5 =0$. In view of the above equation \eqref{Lvlcond2} separates, the compatibility of the separated equation for $U$ with \eqref{spc3} and \eqref{spc5} gives us that $\lambda = \alpha_5^2$, thus we conclude that $\lambda =0$ in this case as well.\\

Case (iii): both of the following conditions hold: 
%This follows from the integrability condition \eqref{LvlCond1}.
\begin{align}\label{spc14}
  \left(\frac{1}{f'}\left( \frac{U''}{U}\right)'\right)'=0.  
\end{align}
\begin{align}\label{spc15}
  \left(\frac{1}{g'}\left( \frac{V''}{V}\right)'\right)'=0.  
\end{align}
The solutions of \eqref{spc14} and \eqref{spc15} are given by
\begin{align}\label{spc16}
    \frac{U''}{U}=\alpha_5 f + \alpha_6
\end{align}
\begin{align}\label{spc17}
    \frac{V''}{V}=(\beta_5 f + \beta_6)
\end{align}
Computing derivatives of the above equations:
\begin{align}\label{18}
&U^{(3)}=(\alpha_5 f + \alpha_6)U'+\alpha_5 f'U \\
&U^{(4)}=2\alpha_5 f'U'+(\alpha_5 f''+(\alpha_5 f + \alpha_6)^2)U \\
&V^{(3)}=(\beta_5 g + \beta_6)V'+\beta_5 g'V \\
&V^{(4)}=2\beta_5 g'V'+(\beta_5 g''+(\beta_5 g + \beta_6)^2)V
\end{align}
With the use of the above derivatives the integrability condition \eqref{LvlCond1} separates to yield the following equations:
\begin{align}\label{spc19}
 &2(\beta_5-\alpha_5) f'U'=((\alpha_5+\beta_5)f''+\alpha_5(\alpha_5+2\beta_5)f^2 \\
 &+(2\alpha_6(\alpha_5+\beta_5)-\alpha)f-3\lambda f^2+\alpha_6^2-\alpha_7)U,  
\end{align}
\begin{align}\label{spc20} 
 &2(\alpha_5-\beta_5) g'V'= ((\alpha_5+\beta_5)g''+\beta_5(\alpha_5+2\beta_5)g^2 \\
 &+(2\beta_6(\alpha_5+\beta_5)+\alpha)g-3\lambda g^2+\alpha_6^2-\alpha_7)V,  
\end{align}
where $\alpha$ is the separation constant. If $\beta_5=\alpha_5$, \eqref{spc19} and \eqref{spc20} imply that 
\begin{align}\label{spc21}
 \lambda=\alpha_5^2, \quad   \beta_5=\alpha_5, \quad \beta_6=-\alpha_6
\end{align}
We conclude that \eqref{spc21} implies that $\phi(u,v)=U(u)V(v)$ defines a separable solution of the Helmholtz equation (See \eqref{SepLvl}.).
If $\beta_5 \neq \alpha_5$, one differentiates \eqref{spc19} and \eqref{spc20} and eliminates all derivatives of $U$ and $V$. One obtains polynomial equations in $f$ and $g$ which imply that $\beta_5^2=\alpha_5^2$. The case $\beta_5=\alpha_5$ has already been considered. The case $\beta_5=-\alpha_5$, yields $3\lambda=-\alpha_5^2$, which is un-physical. This completes the proof of Case (iii).

%\printbibliography

%\end{document}

\end{document}